\documentstyle[11pt,twoside,paspconf]{article}
\begin{document}
%---------------start of defs.-----------------------------------------
%defs.tex
\def\abs#1{\left| #1 \right|}
\def\EE#1{\times 10^{#1}}
\def\gcm{\rm ~g~cm^{-3}}
\def\cm3{\rm ~cm^{-3}}
\def\kms{\rm ~km~s^{-1}}
\def\cms{\rm ~cm~s^{-1}}
\def\ergs{\rm ~ergs~s^{-1}}
\def\isotope#1#2{\hbox{${}^{#1}\rm#2$}}
\def\wl{~\lambda}
\def\wll{~\lambda\lambda}
\def\HI{{\rm H\,I}}
\def\HII{{\rm H\,II}}
\def\HeI{{\rm He\,I}}
\def\HeII{{\rm He\,II}}
\def\HeIII{{\rm He\,III}}
\def\CI{{\rm C\,I}}
\def\CII{{\rm C\,II}}
\def\CIII{{\rm C\,III}}
\def\CIV{{\rm C\,IV}}
\def\NI{{\rm N\,I}}
\def\NII{{\rm N\,II}}
\def\NIII{{\rm N\,III}}
\def\NIV{{\rm N\,IV}}
\def\NV{{\rm N\,V}}
\def\NVI{{\rm N\,VI}}
\def\NVII{{\rm N\,VII}}
\def\OI{{\rm O\,I}}
\def\OII{{\rm O\,II}}
\def\OIII{{\rm O\,III}}
\def\OIV{{\rm O\,IV}}
\def\OV{{\rm O\,V}}
\def\OVI{{\rm O\,VI}}
\def\OVII{{\rm O\,VII}}
\def\OVIII{{\rm O\,VIII}}
\def\CaI{{\rm Ca\,I}}
\def\CaII{{\rm Ca\,II}}
\def\NeI{{\rm Ne\,I}}
\def\NeII{{\rm Ne\,II}}
\def\NeIII{{\rm Ne\,III}}
\def\NeIV{{\rm Ne\,IV}}
\def\NeV{{\rm Ne\,V}}
\def\NaI{{\rm Na\,I}}
\def\NaII{{\rm Na\,II}}
\def\NiI{{\rm Ni\,I}}
\def\NiII{{\rm Ni\,II}}
\def\FeI{{\rm Fe\,I}}
\def\FeII{{\rm Fe\,II}}
\def\FeIII{{\rm Fe\,III}}
\def\FeV{{\rm Fe\,V}}
\def\FeVII{{\rm Fe\,VII}}
\def\CoII{{\rm Co\,II}}
\def\CoIII{{\rm Co\,III}}
\def\ArI{{\rm Ar\,I}}
\def\MgI{{\rm Mg\,I}}
\def\MgII{{\rm Mg\,II}}
\def\SiI{{\rm Si\,I}}
\def\SiII{{\rm Si\,II}}
\def\SiIII{{\rm Si\,III}}
\def\SiIV{{\rm Si\,IV}}
\def\SiVI{{\rm Si\,VI}}
\def\SI{{\rm S\,I}}
\def\SII{{\rm S\,II}}
\def\SIII{{\rm S\,III}}
\def\SIV{{\rm S\,IV}}
\def\SVI{{\rm S\,VI}}
\def\FeI{{\rm Fe\,I}}
\def\FeII{{\rm Fe\,II}}
\def\FeIII{{\rm Fe\,III}}
\def\FeIV{{\rm Fe\,IV}}
\def\FeVII{{\rm Fe\,VII}}
\def\kI{{\rm k\,I}}
\def\kII{{\rm k\,II}}
\def\La{{\rm Ly}\alpha}
\def\Ha{{\rm H}\alpha}
\def\Hb{{\rm H}\beta}
\def\Lya{{\rm Ly}\alpha}
\def\etscale#1{e^{-t/#1^{\rm d}}}
\def\etscaleyr#1{e^{-t/#1\,{\rm yr}}}
\def\sigmaKN{\sigma_{\rm KN}}
\def\ncrit{n_{\rm crit}}
\def\Emax{E_{\rm max}}
\def\chieff{\chi_{\rm eff}^{\phantom{0}}}
\def\chieffi{\chi_{{\rm eff},i}^{\phantom{0}}}
\def\chiion{\chi_{\rm ion}^{\phantom{0}}}
\def\chiioni{\chi_{{\rm ion},i}^{\phantom{0}}}
\def\Gammaion{\Gamma_{\!\rm ion}}
\def\Mcore{M_{\rm core}}
\def\Rcore{R_{\rm core}}
\def\Vcore{V_{\rm core}}
\def\Menv{M_{\rm env}}
\def\Venv{V_{\rm env}}
\def\Vej{V_{\rm ej}}
\def\Vcthou{\left( {\Vcore \over 2000 \rm \,km\,s^{-1}} \right)}
\def\KK{\rm ~K}
\def\Msun{~M_\odot}
\def\Rsun{~R_\odot}
\def\Msunyr{~M_\odot~{\rm yr}^{-1}}
\def\Mdot{\dot M}
\def\M56{M_{\rm ej,56}}
\def\MZA{M_{\rm ZAMS}}
\def\tyr{t_{\rm yr}}
\def\gff{g_{\rm ff}}
\def\Tex{T_{\rm ex}}
\def\math#1{\vskip20true pt\hskip3true cm{#1}}
\def\no{\hang\noindent}
\def\dots{$\ldots$}
\def\etal{{\it et al.}}
\def\ie{{\it i.~e.\ }}
\def\ien{{\it i.~e.~}}
\def\p{\partial}
\def\pp#1{{\partial \over {\partial{#1}}}}
\def\der#1{{d \over {d{#1}}}}
\def\bigskip{\vskip1true cm}
\def\eg{e.~g.~}
\def\prop{\propto}
\def\lsim{\!\!\!\phantom{\le}\smash{\buildrel{}\over
  {\lower2.5dd\hbox{$\buildrel{\lower2dd\hbox{$\displaystyle<$}}\over
   \sim$}}}\,\,}

\def\gsim{\!\!\!\phantom{\ge}\smash{\buildrel{}\over
  {\lower2.5dd\hbox{$\buildrel{\lower2dd\hbox{$\displaystyle>$}}\over
   \sim$}}}\,\,}

%---------------end of defs.-----------------------------------------

\title{The Structure of the Circumstellar Gas of SN 1987A}

\author{Peter Lundqvist}
\affil{Stockholm Observatory, SE-133 36 Saltsj\"obaden, Sweden}
\author{George Sonneborn}
\affil{NASA/Goddard Space Flight Center, Code 681, Greenbelt, MD 20771}

\begin{abstract}
Recent observations of the rings around SN 1987A are discussed and modeled,
with particular emphasis on {\it HST} observations of the inner ring by the 
SINS\altaffilmark{1} team. It is found that the lowest density detected in the 
ring is $\sim (1-2)\EE3 \cm3$. The geometry of the inner ring is constrained by 
its different size in [N II] and [O III]. The implications of this on the 
distance to the supernova are discussed and we find $\lsim 54.2\pm2.2$~kpc, 
which is in agreement with recent RR Lyrae and Cepheid measurements. In 
addition, preliminary results are presented for improved calculations of the
supernova breakout.
\end{abstract}
\altaffiltext{1}{Supernova INtensive Study collaboration (PI: R.~P. Kirshner)}

\section{Introduction}
The rings of SN 1987A have been monitored by an arsenal of ground-based and
space borne telescopes ever since the first detection by {\it IUE} of narrow
circumstellar lines $\sim 70$ days after the explosion (Fransson et al. 1989).
While the structure of the emitting gas was first displayed using {\it CTIO} and 
{\it Las Campanas} instruments (Crotts, Kunkel, \& McCarthy 1989; Crotts, this 
volume) as well as the {\it NTT} at {\it ESO} (Wampler et al. 1990), the 
detailed structure of the rings was revealed only after the installation of 
{\it COSTAR} on {\it HST} (Burrows et al. 1995; Garnavich, this volume). 
Recently, the rings have been observed in a multitude of optical lines using 
{\it STIS} on {\it HST} (Pun et al. 1997), and both UV and IR studies will
follow shortly.

Modeling of the emission lines during the first $\sim 5$ years (Lundqvist \& 
Fransson 1996) has shown that the inner ring consists of gas with a range of 
densities ($\sim 6\EE3 - 3\EE4 \cm3$), and that it is overabundant in helium
and nitrogen compared to normal LMC abundances. These models also show that
emission lines can be used to constrain models of the supernova breakout. In
particular, it is found that the spectrum during the breakout was probably not
very different from that in the models of Ensman \& Burrows (1992), i.e., the
color temperature was in the range $(1.0-1.5)\EE6$~K. However, as is shown 
below, the spectrum is not a simple blackbody.

Our knowledge of the physical conditions in the outer rings is more uncertain.
Panagia et al. (1996) found from nebular analysis that the outer rings had an
electron density of $\sim 8\EE2 \cm3$, at least 2887 days after the outburst,
and that the material in these rings may be less CNO-processed than in the inner
ring. As argued by Crotts, Kunkel, \& Heathcote (1995), and predicted by many 
of the models for the formation of the nebula (e.g., Blondin \& Lundqvist 1993;
Martin \& Arnett 1995; Chevalier \& Dwarkadas 1995; Chevalier, this volume), 
the inner and outer rings may be joined physically. It remains to be seen if 
this is consistent with the different mass loss episodes for the inner and 
outer rings, as proposed by Panagia et al. (1996).

Here we discuss and model more recent data than in Lundqvist \& Fransson (1996).
We focus mainly on {\it HST} observations until late 1996. In addition,
we provide a short discussion about the distance to the supernova, and show 
some recent results from improved models of the supernova breakout.

\section{Observations and Results}

\subsection{{\it HST} images}
\nobreak
A detailed analysis of the structure of the inner ring was made by Plait et
al. (1995) using pre-{\it COSTAR} images. Here we do a similar analysis for
{\it WFPC2} images obtained 2755, 3270 and 3478 days after the outburst. 
While the observations themselves are discussed by Garnavich (this volume; see 
also Lundqvist et al. 1997b), we focus on the interpretation of the 
observations. In particular, we model the angle dependent fading/brightening
of the ring in [N~II] and [O~III], since this can be used to derive the angular 
density distribution of the emitting gas (see Plait et al. 1995). Compared to 
Plait et al. (1995) our spatial resolution is significantly improved, and we 
include also [N~II]. We emphasize that the study of Plait et al. probes gas of 
higher electron density ($n_{\rm e} \gsim 8\EE{3} \cm3$) than our study simply 
because their high-density gas has now cooled and recombined, and does not 
contribute to the emission at our epochs. This means that we also probe gas of 
lower density than Lundqvist \& Fransson (1996) who mainly concentrated on
{\it IUE} data from the first $\sim 1800$~days.

As in Plait et al., we assume that the ring is first heated and ionized by 
the supernova EUV/soft X-ray burst, and then left to recombine and cool. We 
adopt the same model of the burst, i.e., the 500full1 model by Ensman \& 
Burrows (1992), which was successfully used by Lundqvist \& Fransson (1996) 
to model the light curves of the narrow UV lines observed by IUE (Sonneborn 
et al. 1997a). 

We have calculated the evolution of the [N~II] and [O~III] emissivities for 18 
densities ranging from $8.4\EE{2} \cm3$ to $4.2\EE{4} \cm3$, using equidistant 
steps in log(density). The density we have used is the number density of atoms.
The elemental abundances are H: He: C: N: O: Ne: Na: 
Mg: Al: Si: S: Ar: Ca: Fe = 1.0: 0.20: $4.2\EE{-5}$: $1.9\EE{-4}$: 
$1.9\EE{-4}$: $6.2\EE{-5}$: $1.0\EE{-6}$: $1.5\EE{-5}$: $1.2\EE{-6}$: 
$1.7\EE{-5}$: $5.6\EE{-6}$: $3.2\EE{-6}$: $1.1\EE{-6}$: $3.4\EE{-5}$. 
The abundance ratio (C+N+O)/(H+He+Z) is thus $\approx 0.35$ times the solar 
ratio of Anders \& Grevesse (1989). Note also that we assume N/O = 1. 
With these abundances, the electron density is $\approx 1.17$ times the atomic
density when the gas is fully ionized. Abundances and atomic data are discussed
in greater detail in Lundqvist et al. (1997a). (See also \S 2.2)

Figure~1 displays the evolution of the normalized emissivities of 
[N~II] and [O~III] for three of the densities. With increasing density the 
emissivity peaks at earlier times, $t_{\rm peak} \propto n^{-1}$. The results 
shown in Figure~1 are for two cases: one is for an ionization-bounded ring of 
constant density, and the other (henceforth referred to as ``truncated'' 
or ``density-bounded'') takes into account only the innermost 44 \% of the 
H II-region in the ionization-bounded model. A truncated model does not 
necessarily have to be devoid of gas outside the cut, it may just have a sharp 
density drop like in the interacting-winds model (cf. Luo 1991 and Blondin \& 
Lundqvist 1993).

Figure~1 shows that truncation affects strongly the evolution of the 
emissivity from the ring. For example, the maximum [N~II] emission can in the 
ionization-bounded case only be a factor of $\sim 2.5$ higher than immediately 
after the outburst, whereas in the truncated case, arbitrarily large ratios 
are possible. This is because the [N~II] emission in the truncated case 
relies on recombination from N$^{2+}$ and higher states, while in the 
ionization-bounded case, N$^{+}$ is present from the outset.
A similar effect is seen for [O~III], though some O$^{2+}$ is present inside 
the 44 \%-boundary also from the outset (cf. Fig.~1 of Lundqvist \& Fransson
1996). In Figure~1, we have marked our observed epochs by vertical 
dashed lines. (Note that for [O~III] there are only two epochs included.) 

\begin{figure}
\plotfiddle{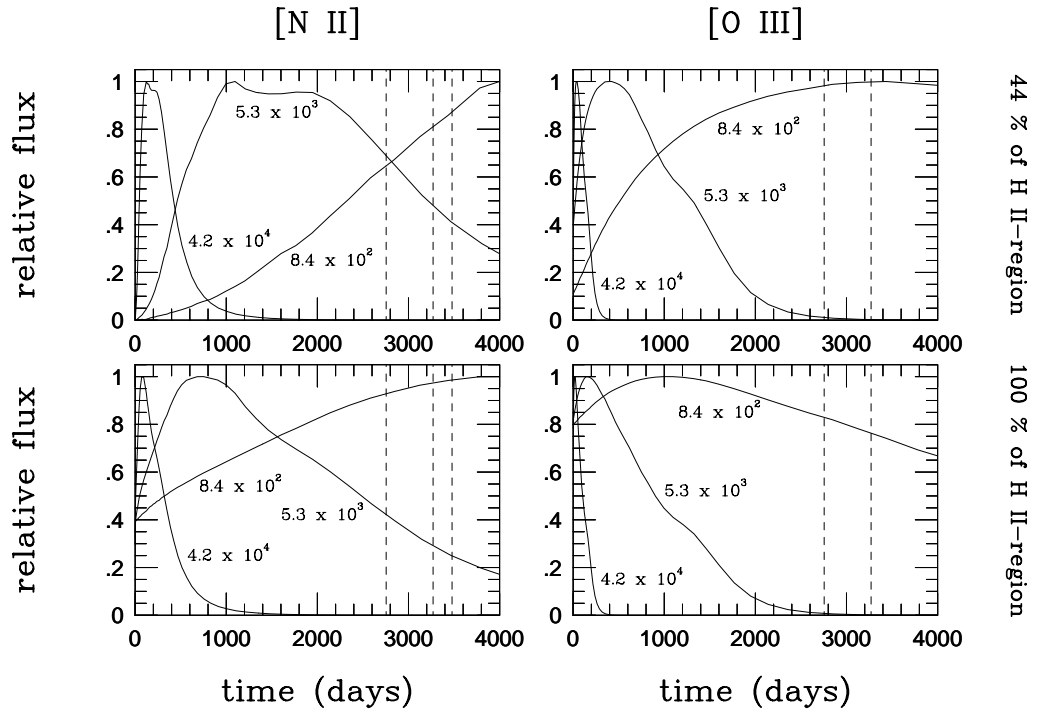}{8cm}{0}{70}{70}{-220}{-260}
\caption 
{\scriptsize{Evolution of relative emissivities of [N~II] and [O~III] from the 
inner ring for three single-density models: $8.4\EE{2} \cm3$, $5.3\EE{3} \cm3$
and $4.2\EE{4} \cm3$. The highest-density model peaks first. The bottom row of 
panels ($100 \%$) is for an ionization-bounded ring, while in the top row, only 
the $44 \%$ innermost region of the H II-region is included. (Such models are 
referred to as ``truncated'' in the text.) The observed epochs are marked with 
dashed vertical lines. No compensation for light travel times have been 
included in the figure.}}
\end{figure}

To compare Figure~1 with the observed position-angle (PA) dependent
fading/brightening between epochs, light travel times have to be included 
(e.g., Dwek \& Felten 1992). We have assumed a tilt angle of the ring 
of $43$~degrees and a ring radius of $6.3\EE{17}$ cm (e.g., Sonneborn et al. 
1997a). This gives minimum and maximum shifts between the observed epoch and 
the epoch of emission of $\sim 77$~days (for PA $\sim 0$~degrees) 
and $\sim 409$~days (for PA $\sim 180$~degrees), respectively. As these numbers 
are much smaller than the number of days since the explosion at the observed 
epochs, correction for light travel times is less important in the current 
study than in Plait et al. (1995).

\begin{figure}
\plotfiddle{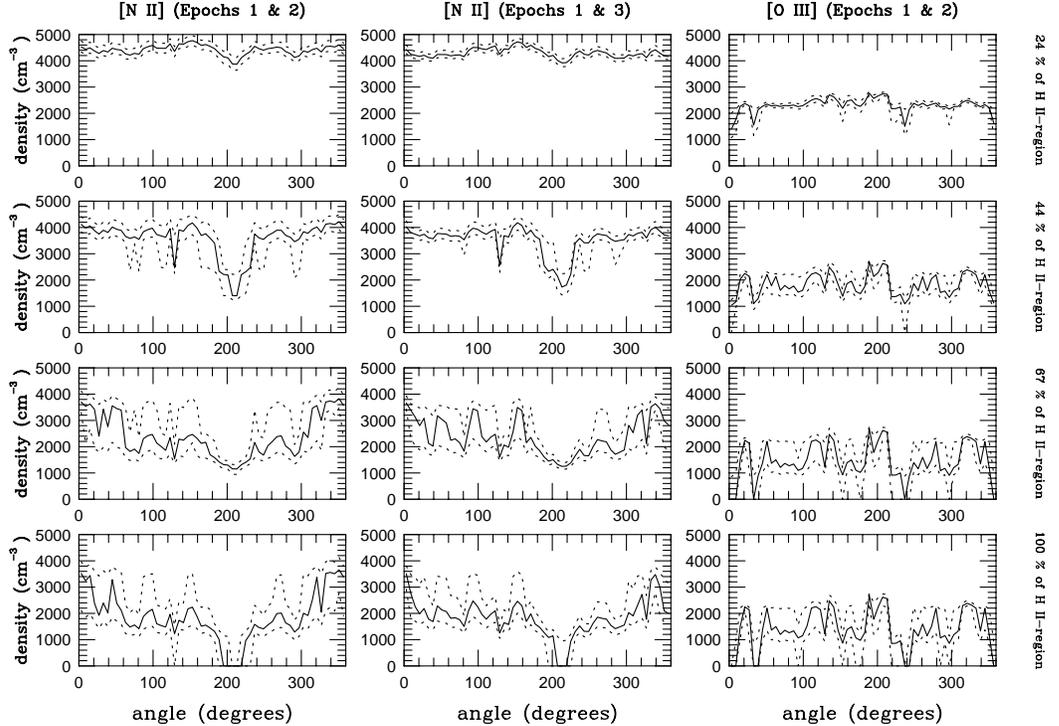}{8cm}{0}{80}{80}{-255}{-315}
\caption 
{\scriptsize{Modeled mean density of the inner ring as a function of position 
angle for four degrees of truncation of the ring. The bottom row of panels is 
for an ionization-bounded ring, while going topwards, the ring has been 
truncated closer and closer to its inner edge. Solid lines are for the measured
ratios of fluxes between the epochs listed at the top of each column, whereas 
dotted lines are for a $5 \%$ uncertainty of these ratios. Light travel times 
have been included.}}
\end{figure}

The density corresponding to the fading/rebrightening of [N~II] and [O~III]
between the observed epochs described in Lundqvist et al. (1997b) is shown in
Figure~2, where we have also added results for two other locations of 
the truncation ($24 \%$ and $67 \%$). The solid lines in Figure~2
correspond to the observed ratios, whereas the dotted lines correspond to 
a $\pm 5\%$ uncertainty of these ratios. In our analysis there are also 
systematic errors due to uncertainties in atomic data as well as approximations
in the photoionization code. However, we believe that the dominant uncertainty
affecting the estimated density is the degree of truncation. It should be 
emphasized that our analysis, like the one of Plait et al. (1995), only gives
a mean density for each position angle. In reality, this is an average of both
higher and lower values. 

Figure~2 shows that the estimated density is lower in ionization-bounded
models ($\sim 2\EE{3} \cm3$ for [N~II] and $(1-2)\EE{3} \cm3$ for [O~III])
than in truncated, especially in the case of [N~II]. This is not surprising 
since truncated models rely on recombination to emit the optical lines. The 
time scale for this is $t_{\rm rec} \propto n_{\rm e}^{-1}$. Because the highest
degree of ionization at the outset occurs close to the inner edge of the ring,
higher densities are needed to reach N$^{+}$ at the observed epoch the more 
the ring is truncated.  We cannot determine from Figure~2 alone the degree of 
truncation, though {\it it seems evident that at least some truncation is
necessary at PA $200 - 210$~degrees to explain the rebrightening of [N~II]} 
at that position angle reported by Garnavich \& Kirshner (1996); 
it is only in truncated models that one gets a sufficient 
increase in emissivity between the observed epochs. {\it The fact that 
truncated models are able to explain the rebrightening of [N~II] also means 
that no additional ionizing source is needed}, which is fully consistent with
the finding of Lundqvist et al. (1997a) that the observed X-ray emission 
(Hasinger, Aschenbach, \& Tr\"umper 1996; Hasinger, this volume) is too weak
to reionize the ring. We emphasize that if the ring is truncated also at 
other position angles, the degree of truncation may vary around the ring. 
Different panels of Figure~2 may therefore apply at different
position angles of the ring. We will, however, in the following assume that
the same truncation applies at all position angles. We have also disregarded 
the effect that different position angles sample different path lengths through 
the ring along the line of sight. This is a simplification since observations
at different position angles may bias sligthly different physical conditions.

\subsection{{\it HST} spectra}
\nobreak
\begin{figure}
\plotfiddle{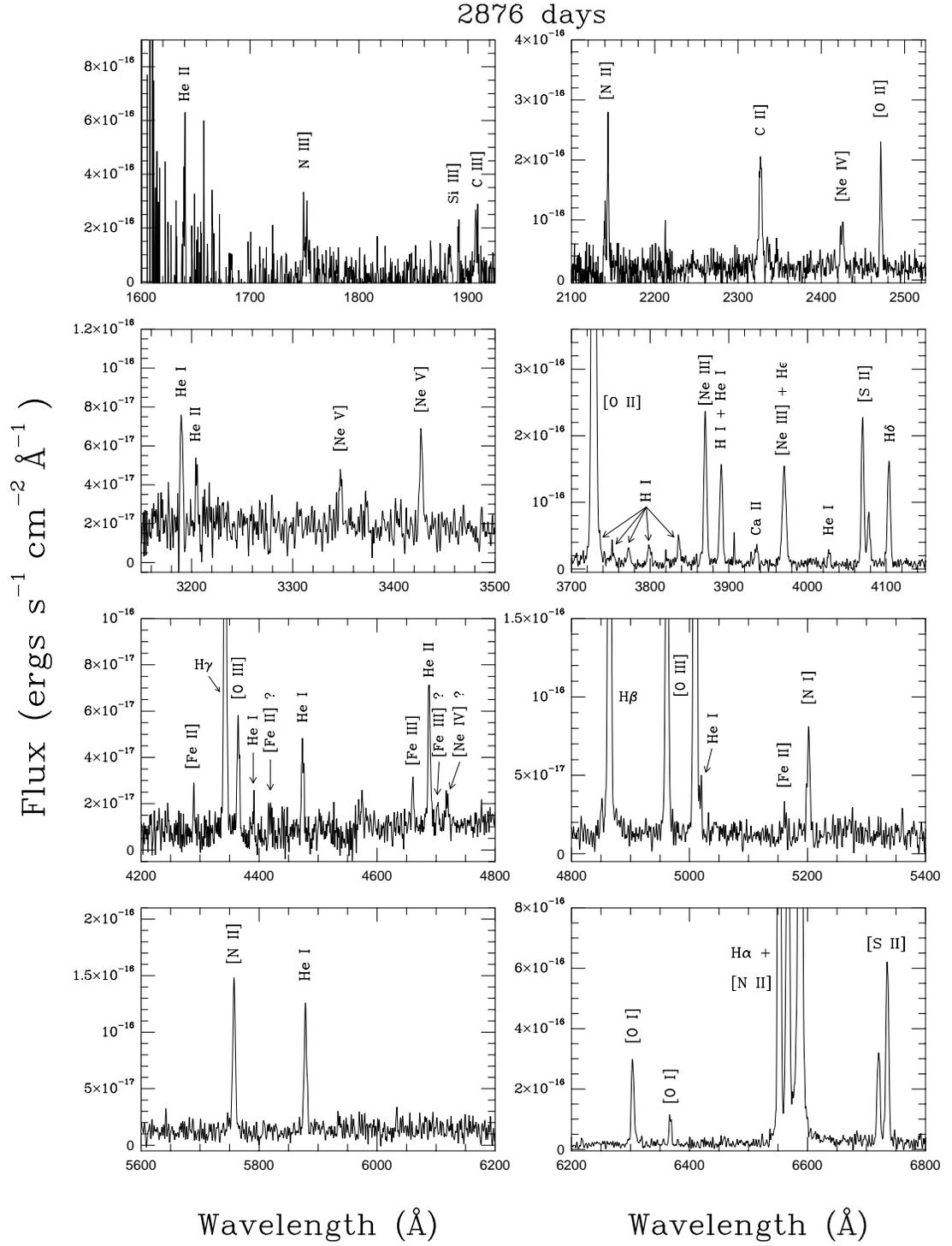}{17cm}{0}{80}{80}{-255}{-50}
\caption 
{\scriptsize{{\it HST/FOS} spectrum of the inner ring at PA 300~degrees at 
2876 days. Both clear and tentative line identifications have been marked. The
spectrum has not been dereddened.}}
\end{figure}
Spectra of the inner ring have been taken within the SINS collaboration at the 
four epochs 1864, 2228, 2876 and 3262 days. (Each of these epochs is actually 
a ``mean epoch'' weighted by exposure time.) A full description and analysis of 
the spectra is done in Lundqvist et al. (1997a), and a nebular analysis is 
provided by Panagia, Scuderi, \& Gilmozzi (this volume). Most of the lines
identified by Lundqvist et al. were identified prior to the {\it HST} 
observations (Wang 1991; Cumming 1994; Sonneborn et al. 1997a), but many are 
reported here, and by Lundqvist et al. (1997a), for the first time, especially 
lines in the UV due to the higher sensitivity of the {\it HST} compared to the 
{\it IUE}.
Previously unidentified lines are: $\HeI$~$\lambda$3188, $\HeI$~$\lambda$4388,
$\HeII$ $\lambda$2734, $\HeII$ $\lambda$3203, $\CII$] $\lambda$2326,
[$\NI$] $\lambda\lambda$5198, 5200,
[$\NII$] $\lambda\lambda$2139, 2143, [$\NII$] $\lambda$3063,
[$\OII$] $\lambda$2470, [$\NeIV$] $\lambda\lambda$2422, 2424,
[$\NeV$] $\lambda$2975, [$\NeV$] $\lambda\lambda$3346, 3426,
Mg I] $\lambda$4792, $\SiIII$ $\lambda\lambda$1883, 1892,
[$\FeII$] $\lambda\lambda$4287, 4416 and [$\FeIII$] $\lambda$4702.
As an example of the spectra, we show in Figure~3 selected parts of the day 
2876 spectrum. This observation was centered on the brightest part of the
ring, i.e., that at PA 300~degrees.

The dominant lines at all four epochs are [$\NII$]~$\lambda\lambda$6548, 6583.
Combined with [$\NII$]~$\lambda$5755, these give a temperature 
of $\sim 1.0\EE4$~K at all epochs. (We have assumed $E(B-V) = 0.16$, in 
accordance with Sonneborn et al., 1997a). The temperature derived from
the [N~II] UV doublet, $I_{\lambda\lambda2139, 2143}$/$I_{\lambda5755}$ 
is $30-50$\% higher. This is accounted for by the fact that the UV doublet has
a higher excitation energy than [N~II]~$\lambda\lambda$6548, 6583, which means
that the UV doublet is primarily emitted in regions with higher temperature 
than those emitting the [$\NII$]~$\lambda\lambda$6548, 6583 lines. The 
temperature derived from [$\OIII$]~$\lambda\lambda$4959, 5007 and 
[$\OIII$]~$\lambda$4363 is even higher, ($2.3-3.0) \times 10^4$~K.

Densities are most easily obtained from [O~II] and [S~II] line ratios. In
general, the [S~II]~$\lambda\lambda$6716, 6731 lines yield $\sim 1.0 \EE4 \cm3$,
whereas the more highly excited [S~II]~$\lambda\lambda$4069, 4076 lines give
a somewhat lower density, and indicate a [S~II] temperature of $\lsim 10^4$~K
after $\sim 2228$~days. The [O~II] densities are lower 
(${\rm few} \times 10^3 \cm3$) and indicate temperatures 
approaching $2\EE4$~K.

The abundances estimated from modeling of the spectra (Lundqvist et al. 1997a)
are close to those listed in \S 2.1. Apart from the elements discussed in
detail by Lundqvist \& Fransson (1996), neon appears to be consistent with its
normal LMC abundance, while silicon is underabundant by a factor of $\sim 2$.
The low silicon abundance could be due to depletion caused by grain formation
in the red supergiant wind of the progenitor prior to the explosion. An argument
along the same line was put forward by Borkowski, Blondin, \& McCray (1997) for
a low iron abundance. Unfortunately, the iron lines detected by {\it HST} are 
too weak, and atomic data for iron is rather poor, so we cannot confirm their 
low iron abundance. However, an analysis of ground-based data from {\it ESO/NTT} 
is underway (Cumming \& Lundqvist, in preparation).
\begin{figure}
\plotfiddle{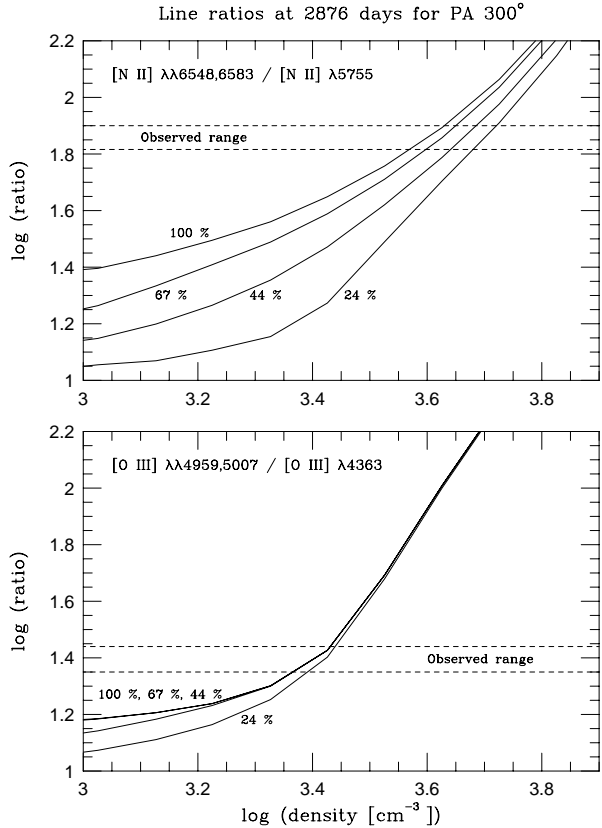}{11cm}{0}{45}{45}{-150}{-15}
\caption 
{\scriptsize{[N~II] and [O~III] line ratios for PA $300$~degrees at 2876 days 
as functions of density for the four degrees of truncation described in the 
text (solid lines). Dashed lines show the observed range of the {\it HST/FOS} 
observations of Lundqvist et al. (1997a) at the same epoch. Note that the 
[N~II] density is a factor of $\sim 2$ higher than that of [O~III], and that
neither of these densities are as sensitive to truncation as the densities 
in Figure~2.}}
\end{figure}

\section{Discussion}
\subsection{Consistency check of density from spectra and images}
\nobreak
To distinguish between the densities in Figure~2, and thus the importance of 
truncation, we have compared our findings in \S 2.1 to the densities derived
by Lundqvist et al. (1997a). We concentrate on their observation at 2876 days 
(see Fig.~3) since this is the closest in time to the epochs of the images. Just
as for 2876 days, the spectrum at 3262 days was taken of the brightest part of 
the ring. For both epochs Lundqvist et al. derive [N~II] and [O~III] 
temperatures of $\sim 1.0\EE{4}$~K, and $\sim 2.7\EE{4}$~K, respectively. In 
Figure~4, we compare their dereddened intensity ratios for [N~II],
$I_{\lambda\lambda6548, 6583}$/$I_{\lambda5755}$, and for [O~III],
$I_{\lambda\lambda4959, 5007}$/$I_{\lambda4363}$, with ratios 
obtained from the truncated and ionization-bounded single-density models 
discussed in Figures~1 and 2. For the observed ratios we have assumed an 
uncertainty of $\pm 10 \%$. 

Figure~4 shows that the ratio of the [O~III] lines favors a mean density 
of $(2.5 \pm 0.3)\EE{3} \cm3$, rather independent on the amount of 
truncation. This is consistent with the [O~III] density at PA $300$~degrees
in Figure~2, especially in the case of the $24 \%$-model. Because we cannot 
expect a mean density found from the fading of the [O~III]~$\lambda5007$ 
emission to be exactly the same as that found from the 
$I_{\lambda\lambda4959, 5007}$/$I_{\lambda4363}$ ratio, none of 
the other models in Figure~2 can, however, be excluded by the [O~III] lines. 
The situation is clearer for the [N~II] lines. Here the temperature sensitive 
ratio gives a density of $(5.0\pm0.3)\EE{3} \cm3$ for the $24 \%$-model, 
decreasing down to $(4.0\pm0.3)\EE{3} \cm3$ for the ionization-bounded model.
Comparing with the first column of panels in Figure~4, which corresponds best 
in time to the {\it FOS} observations, best agreement is again obtained for 
the $24 \%$-model, while both the $67 \%$ and ionization-bounded models
can probably be excluded. Assuming that the degree of truncation be the same 
around the ring, Figures~2 and 4 show that the average density of the 
[N~II]-emitting gas is probably in excess of $\sim 4.0\EE{3} \cm3$, while that
of the [O~III]-emitting gas is $\sim (2.0-2.5)\EE{3} \cm3$.

The slightly higher density we find for [N~II] than [O~III] is in accordance
with the fact that [S~II]~$\lambda\lambda$6716,~6731 comes from gas of even 
higher density; at 2876 days  Lundqvist et al. (1997a) find that the [S~II] 
density is $\sim 9\EE{3} \cm3$. On the other hand, a density derived from
[O~II]~$\lambda\lambda3726,~3729$ should be lower than that characterizing
the [N~II] emission, despite the same ionization stage. This is because 
the [O~II] lines have higher excitation energies, and are therefore biased 
towards hotter gas, i.e., low-density gas that has not cooled down to $10^4$~K. 
Lundqvist et al. (1997a) find a density from [O~II] which is similar to ours
for the [O~III] lines, as well as that estimated by Panagia et al. (1996)
from [O~II]~$\lambda\lambda3726,~3729$ at 2887 days. 

Because our images show a bias for the [N~II] emission towards the inner edge 
of the [O~III] ring, we speculate that the [S~II] emission should spatially
trace the [N~II] emission, while the [O~II]~$\lambda\lambda$3726,~3729 emission
probably correlates more with the [O~III] emission. The 
[O~II]~$\lambda\lambda$7320-7330 multiplet, however, could be more concentrated
to the [N~II] emission, since it has a lower excitation energy than 
[O~II]~$\lambda\lambda$3726,~3729. Emission in [O~I]~$\lambda\lambda$6300,~6364 
may trace even higher densities than [S~II]. Observations using {\it STIS} 
should soon sort this out.

\subsection{Geometry of the inner ring}
\nobreak
\begin{figure}
\plotfiddle{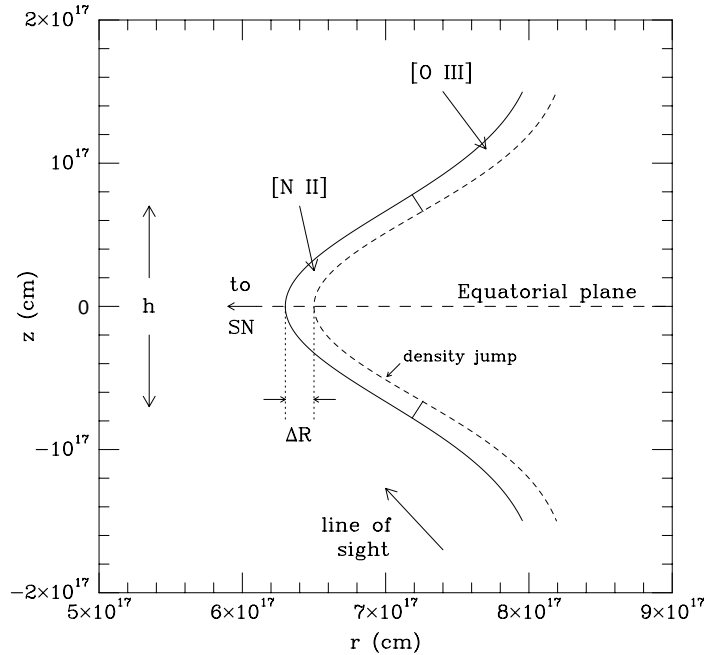}{9cm}{0}{50}{50}{-160}{-70}
\caption 
{\scriptsize{Sketched cross section of the ring in its equatorial plane. The 
figure is only intended to show the rough properties of the ring rather than 
its exact geometry. As indicated by the arrow showing the line of sight, 
the cross section applies to PA $0$~degrees. The [N~II] emission comes from
close to the equatorial plane, whereas [O~III] comes from regions more distant
from the supernova, as well as further from the equatorial plane.
The boundary between these regions is indicated by a solid line, though there
is most likely a more gradual decline in density with increasing $\abs{z}$.
The figure only shows the structure for $\abs{z} < 1.5\EE{17}$~cm since there
is no information from the HST images how the ring connects to the outer rings
of the supernova. The label ``density jump'' refers to the fact that 
the [N~II]-emitting gas appears to be density-bounded rather than 
ionization-bounded. The high-density gas observed by {\it IUE} ($n_{\rm e} \sim
{\rm few} \times 10^4 \cm3$) is probably correlated with the [\NII] region
in the figure. The parameters $\Delta$$R$ and $h$ are described in the text.}}
\end{figure}
We now discuss a viable geometrical model to explain the observed structure 
of the emitting gas in the inner ring. In particular, the model must explain 
why the observed [N~II] emission comes from a region inside the [O~III] region,
and why it is thinner (Garnavich, this volume; \S 3.3 below).
Because we have found that $n_{\rm [N~II]} > n_{\rm [O~III]}$
this means that low-density gas is situated, on average, 
outside that of higher density. From the analysis in Lundqvist et al. (1997a),
the most consistent explanation is that $R_{\rm in}$, the inner radius of the
observed ring, is different for the emission in [N~II] and [O~III]. If this is 
correct, it is natural to assume that there is a continuous change in density
from the [N~II] region to the [O~III] region, and that the two regions are 
joined physically. It is also important that we explain why the observed
thickness of the [N~II] gas appears to be roughly the same at the ring's 
semiminor and semimajor axes. This cannot be done with a ring which 
has $h \gg \Delta$$R_{\rm [N~II]}$ (where $h$ is the total extent of the ring
perpendicular to the intrinsic plane of the ring and $\Delta$$R_{\rm [N~II]}$
the radial thickness of the [N~II] ring in the same plane), unless the inner 
surface of the ring has a radius which changes significantly with distance 
above and below the plane of the ring. 

The situation envisaged is thus similar to the ionized-on-the-inside-only 
crescent discussed in Plait et al. (1995), only that in our case we do not 
argue for ionization marking the boundary between emitting and non-emitting 
gas. In our model the nebula observed by {\it HST} after $\sim 2755$ days is
density-bounded.

Figure~5 shows a cross section of the ring in this model. The particular part
of the ring we have chosen to show is at PA $0$~degrees. The model is the same 
for other position angles, only that the line of sight will be different.
Due to the brightness variation with position angle, it is obvious that the
ring is not axisymmetric, so Figure~5 is only meant to be a rough 
representation of the real ring. In particular, $\Delta$$R$, $h$ 
and $R_{\rm in}$ are all likely to vary with position angle. In addition, 
we do not know the exact curvature of the ring as a function of $z$, i.e., the
distance perpendicular to the plane of the ring. However, because of the 
rather well-determined tilt angle of the ring, the curvature cannot be very 
different from that in Figure~5 to obtain [N~II] on the inside of the [O~III]
emission in the observed image. 

Another benefit from our model is that it can account for the fact that both
the [N~II] and [O~III] emission extends over a large 
radius, $\Delta$$R_{\rm obs}$, corresponding 
to $\Delta$$R_{\rm obs} / R_{\rm in} \sim 0.1 - 0.2$, and that it does so for
all position angles. For a ring with no curvature with increasing $\abs{z}$,
this cannot be the case for PA $90$ and $270$~degrees, especially 
if the ring has a filling factor close to unity. With the model in 
Figure~5 there is no need for a low filling factor in the {\it emitting}
region; it is only in {\it projection} the filling factor is low for a given 
line of sight. 

Our model predicts that the [N~II] emission will eventually venture into
the region now emitting [O~III], thus increasing in observed size. At the same 
time, the high-density gas closest to the supernova will cool and fade in 
[N~II]. The [O~III] nebula may also increase in size if there are regions 
of lower density outside the present nebula.

It is interesting that our geometrical model (Fig.~5) bears resemblance with 
the {\it structure} close to the equatorial plane in the models of Blondin \& 
Lundqvist (1993) and Martin \& Arnett (1995). In both simulations, 
the structure curves out toward larger $r$ with increasing $\abs{z}$. 
Ground-based (e.g., Crotts, this volume) and deeper {\it HST} images are needed
to compare modeled and observed structures at large $z$. However, as noted by
Lundqvist \& Fransson (1996), we already know that neither the models of
Blondin \& Lundqvist nor those of Martin \& Arnett are likely to adequately
model the structure as far above and below the equatorial plane as the outer 
rings.

Our constraints on the geometry and density of the ring are important also for 
the modeling of the forthcoming ejecta/ring interaction (Borkowski et al. 1997).
The most obvious change to the model of Borkowski et al. 
is that there will be no shock entering the ring from ``behind'', i.e.,
the side of the ring opposite to that facing the supernova, unless the ring 
is broken up in the third dimension. With the relative distribution of 
components with different densities discussed in Lundqvist et al. (1997a,b), 
better predictions of the line emission from the ejecta/ring nebula can be
made. Conversely, our model of the nebula can be tested as the ejecta start 
interacting with different parts of the ring at different times.

\subsection{Distance to the supernova}
\nobreak
Panagia et al. (1991) devised a simple way to estimate the distance to SN 1987A
using the {\it IUE} light curves of the narrow lines, in combination with 
{\it HST} imaging observations. The method assumes a perfect match between the
geometry of the UV-line emitting gas around day 80 to 400 and that emitting 
[O~III] $\sim 1000$~days later. It is also thought that the gas does not have
to recombine before emitting the UV-lines, and the ring is assumed to be 
intrinsically circular. Later, Gould 
\begin{figure}
\plotfiddle{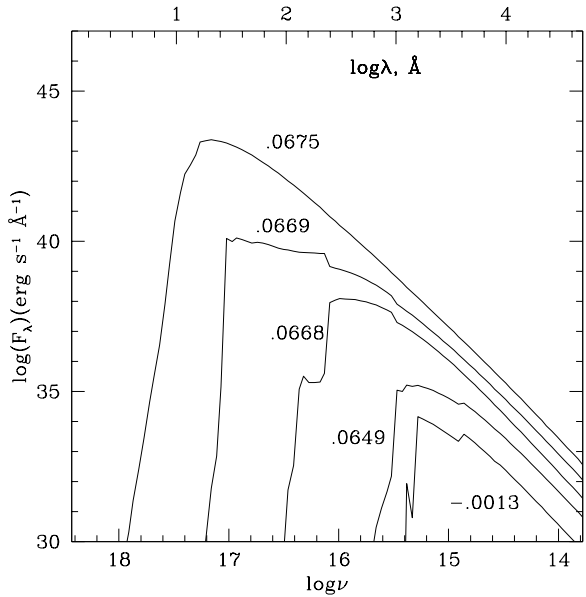}{7cm}{0}{40}{40}{-130}{-55}
\caption 
{\scriptsize{Spectral flux in observer's frame at shock breakout. The progenitor
model has a mixed composition and is from Saio, Nomoto, \& Kato (1988) and
Shigeyama, Nomoto, \& Hashimoto (1988). The mass of the star is $16.3 \Msun$ 
and its radius is $3.4\EE{12}$~cm. The mass cut is at $1.6 \Msun$ and the 
explosion energy is $1.3\EE{51}$~ergs. The curves are labeled by retarded time
in days.}}
\end{figure}
\begin{figure}
\plotfiddle{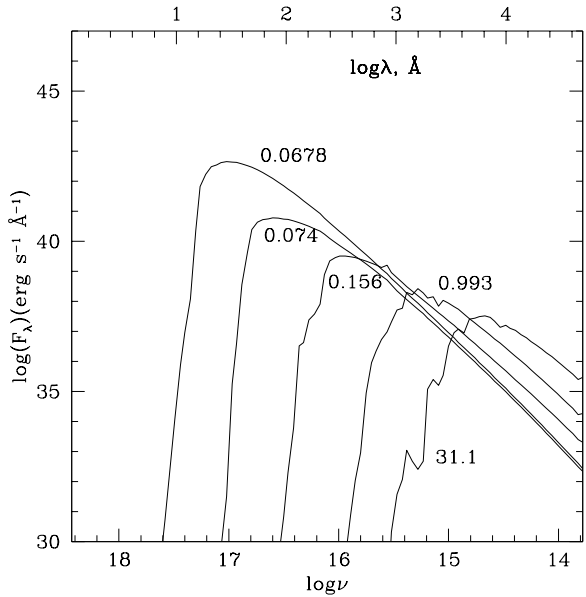}{7cm}{0}{40}{40}{-130}{-55}
\caption
{\scriptsize{Same as Fig.~6, but after shock breakout.}}
\end{figure}
(1995) used a similar approach including a more sophisticated 
statistical treatment. Both studies argued for rather small errors, but did 
not agree on the inferred distance ($51.2\pm3.1$~kpc 
and $\lsim 46.77\pm0.76$~kpc, respectively). Recently, both groups have 
reevaluated their distances (Panagia, Gilmozzi, \& Kirshner, this volume; 
Gould \& Uza 1997, preprint) now obtaining $51.5\pm1.2$~kpc 
and $\lsim 48.76\pm1.13$~kpc, respectively, still inconsistent with each other.
Using the same method Sonneborn et al. (1997a) obtained $48.6\pm2.2$~kpc,
assuming the [O~III] angular diameter of Plait et al. (1995).

It is obvious that the real error of these estimates could be arbitrarily large
simply because the gas once dominating the UV-line emission has a high density
and faded before the first {\it HST} images were taken. To minimize the error
one should measure the angular size of the ring in lines tracing the same gas
as that emitting the UV lines during the first $\sim 400$~days. 
Obviously, [N~II], and especially [O~III], are not ideal in that sense, at least
not at present epochs. Instead, one should observe lines emitted by ions that
have recombined further. Both [S~II] and [O~I] should be better, though [S~II] 
may only be marginally better than the [O~III] observations of Plait et al. 
(1995), which indicated emission from gas of the same density as [S~II] 
indicates today. [O~I] observations have now been made using {\it STIS} (Pun
et al. 1997; Sonneborn et al. 1997b), and we will include this line in future 
analyses.

Meanwhile, our geometrical model in Figure~5, together with the models of
Lundqvist \& Fransson (1996), suggest that the innermost region of the [N~II] 
region is likely to give a reasonable estimate of where the first UV emission 
lines came from. The time of turn on of the UV lines, $t_{\rm min}$, is
therefore related to $R_{\rm in}$ of [N~II], while the time when the UV lines
peak, $t_{\rm max}$, depends on how far from the equatorial plane the 
high-density gas reaches, as well as the off-equatorial geometry of the ring.
The most conservative assumption is that $t_{\rm max}$ corresponds to a radius
which is $\gsim R_{\rm in}$ of [N~II]. From the {\it HST} images in \S 2.1 we 
find that the angular extent corresponding to $R_{\rm in}$ of [N~II] could be as 
low as $\sim 770\pm20$~mas. We combine this with $t_{\rm min} = 84\pm4$~days
and $t_{\rm max} = 399\pm15$~days from Sonneborn et al. (1997a), and obtain a
distance to the supernova which is $\lsim 54.2\pm2.2$~kpc. While this is
only an upper limit, it is consistent with recent estimates of RR-Lyraes and
Cepheids (Reid 1997; Gratton et al. 1997; Feast \& Catchpole 1997), which
all indicate values around our limit. We emphasize that a model assuming an
infinitesimally thin ring combined with published mean values of the angular 
extent of the ring can give unreliable estimates.

\subsection{What is next?}
\nobreak
Future progress in our knowledge about the rings, and the circumstellar nebula
in general, relies on continued monitoring, especially with {\it HST}, as well
as more refined modeling. The most problematic point on the modeling side is 
that there is no convincing model for the formation of the outer rings. 
Hopefully this meeting has sparked some new ideas about that. The model of
Chevalier \& Dwarkadas (1995; see also Chevalier, this volume) appears to be
on the right track, and maybe we can learn something from the rings around
the star Sher \#25 and other blue stars with rings (Chu, Brandner, \& Grebel, 
this volume). 

More detailed analysis of the emission lines from the rings requires more 
detailed calculations of the ionizing spectrum at shock breakout.
Such calculations are in progress using a multigroup method and accurate
expansion opacities (Blinnikov et al. 1997; see also Nomoto, Blinnikov, \&
Iwamoto, this volume). An example of these calculations is shown in Figures~6 
and 7, where also the parameters of the model are given. Although the 
effective temperature is close to that in the models of Ensman \& Burrows 
(1992), the spectrum is different from a superposition of blackbody spectra. 

Finally, progress in our understanding of the rings also requires ground-based 
observations to pick up weak lines neccessary for accurate abundance analyses,
as well as to determine the velocity field (e.g., Crotts, this volume).

\vskip0.7cm
\acknowledgments
We are grateful to the SINS team for collaboration. P.L. wishes to thank
Oleg Bartunov, Sergei Blinnikov, Peter Challis, Robert Cumming, Peter Garnavich
and Garrelt Mellema for collaboration and discussions. This research
was supported by the Swedish Natural Science Research Council, the Swedish
National Space Board and the Wenner-Gren Center Foundation for Scientific
Research.

\end{document}